\documentclass{article}

\usepackage{arxiv}

\usepackage[utf8]{inputenc}

\usepackage[T1]{fontenc}

\usepackage{hyperref,url,booktabs,amsmath,amsfonts,nicefrac,microtype,graphicx,natbib,doi,adjustbox}

\title{Physics-Guided Deep Metric Learning with Continuous Time Embeddings for Open-World Radar Pulse De-Interleaving}

\author{
	\href{https://orcid.org/0009-0007-3284-3423}
	{\includegraphics[scale=0.06]{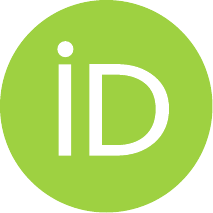}\hspace{1mm}}
	Dr.\ Vikas Agnihotri \\
	Independent Researcher \\
	India \\
	\texttt{\href{mailto:er.vikasagnihotri@gmail.com}{er.vikasagnihotri@gmail.com}}
	\And
	Jasleen Kaur \\
	National Institute of Technology, Rourkela \\
	India \\
	\texttt{\href{mailto:jasstronger@gmail.com}{jasstronger@gmail.com}}
}

\renewcommand{\shorttitle}{Continuous Time Embeddings for Open-World Radar Pulse De-interleaving}

\begin{document}

\maketitle

\begin{abstract}

Radar pulse de-interleaving is a foundational Electronic Support Measures (ESM) task that aims to separate chronologically interleaved pulse streams from multiple non-cooperative transmitters under unknown emitter cardinality in dense, contested electromagnetic environments. Classical histogram transforms and closed-world deep classifiers degrade under severe pulse loss, agile Pulse Repetition Interval (PRI) modulation, and spurious clutter. In this paper, we systematically characterise continuous temporal representations and physics-guided model selection in deep metric learning for open-world radar de-interleaving. Building on the transformer-based metric-learning framework for open-world deinterleaving introduced by Gunn et al. \cite{gunn2025radar}, we introduce a continuous Time-of-Arrival (ToA) sinusoidal positional encoding that directly models physical inter-pulse durations rather than ordinal token indices, a design choice that contrasts with \cite{gunn2025radar}, who found ordinal positional encodings provided no benefit and omitted them entirely. Neural network parameters are optimised solely via Supervised Contrastive (SupCon) learning, while scale-aware physical domain priors based on PRI Consistency and Angle-of-Arrival (AoA) continuity serve as physics-guided validation and checkpoint-selection criteria operating on unsupervised HDBSCAN cluster assignments. Across multi-seed ablations and full open-world benchmarking on the Turing Synthetic Radar Dataset (TSRD) \cite{gunn2026turing} (1,849 test windows observing up to 110 transmitters), our investigation establishes three primary findings: (1) Representation: Continuous physical ToA positional encoding yields a +133.4\% relative F1 improvement (0.8744 vs. 0.3747) and +87.5\% V-measure improvement (0.8484 vs. 0.4526) over discrete ordinal index encodings across three independent random seeds (Table IV); (2) Conditional Physics Priors: On the full open-world TSRD test set, the unconstrained baseline M1 achieves the highest Hungarian F1 (0.5813) and CRR (0.8296). Physics-guided configurations M3 and M4 (AoA-guided, identical by epoch selection) achieve F1 = 0.5790 with substantially lower cross-seed standard deviation ($\pm$0.0013 vs. $\pm$0.0063 for M1). On controlled synthetic streams, PRI-guided selection (M2) provides modest F1 advantages under moderate clutter and pulse loss (up to +0.0125 F1 at 20\% pulse loss), but M1 matches or surpasses M2 at 30\% pulse loss and under combined harsh noise; (3) Systems Implication: End-to-end latency profiling demonstrates sustained throughput exceeding 15,100 PDWs/s (16.85 ms at W = 256), revealing that downstream density clustering (HDBSCAN, 14.85 ms) accounts for 88.1\% of total pipeline latency while the neural encoder pass requires only 2.00 ms.

\end{abstract}


\section{Introduction}

Electronic Support Measures (ESM) and Electronic Intelligence (ELINT) systems play a critical role in modern electronic warfare by intercepting and analysing electromagnetic emissions from non-cooperative radar transmitters \citep{wiley2006elint}. The first and most critical stage of ESM signal processing is radar pulse de-interleaving---the process of sorting a chronologically interleaved stream of received pulses into distinct subsets corresponding to their respective source emitters. Accurate de-interleaving enables downstream radar signal analysis tasks including target characterisation, threat assessment, and platform recognition \citep{agnihotri2019effect}, \citep{agnihotri2019extraction}, \citep{agnihotri2020automatic}.

Once sorted, individual pulse trains can be used for emitter identification, location tracking, threat assessment, and electronic countermeasure generation. Intercepted pulses are typically represented as high-dimensional tabular records known as Pulse Description Words (PDWs). A standard PDW includes Time-of-Arrival (ToA), Radio Frequency (RF), Pulse Width (PW), Angle-of-Arrival (AoA), and Pulse Amplitude (PA).

Modern dense radar environments pose substantial challenges for classical algorithms like Cumulative Difference Histograms \cite{genccol2017improvements} or standard PRI transforms \cite{wang2022parametric}). These are mainly due to three obstacles: agile pulse modulation (such as stagger or jitter modulation schemes) that break periodic assumptions, severe signal degradation consisting of missing pulses (pulse loss) and spurious background pulses (clutter injection), and significant spatial-temporal overlaps in dense emitter environments.

Deep learning treatments usually rely on recurrent neural networks or standard sequence-to-sequence transformers \cite{vaswani2017attention}. Gunn et al. \cite{gunn2025radar} established a transformer-plus-metric-learning pipeline for exactly this open-world setting, using triplet loss and HDBSCAN clustering with no positional encoding at all, reporting that ordinal positional encodings produced a small negative effect in preliminary experiments. This leaves open whether a continuous, physically-grounded encoding of elapsed time — rather than ordinal position — can succeed where ordinal encoding failed. Additionally, closed-world supervised classifiers inherently fail to generalise across unknown and dynamic emitter counts. Furthermore, applying physical domain priors directly to normalised feature representations distorts continuous physical scales, demanding a principled, scale-aware formulation.

Rather than asserting universal dominance of physical domain priors across all operating conditions, this paper provides a rigorous, empirical characterisation of when and why physical domain priors assist versus impair metric representation learning. Specifically, we delineate three core contributions:

\begin{itemize}

\item Continuous Physical Time Representation: We design and evaluate a continuous Time-of-Arrival sinusoidal positional encoding that maps raw microsecond timestamps directly into attention query-key space, demonstrating dramatic improvements (+133.4\% relative F1, +87.5\% V-measure, +156.7\% CRR) over discrete index orderings across multiple independent random initialisations.

\item Characterisation of Conditional Physics-Guided Model Selection: Through extensive sweeps across 200 synthetic streams per noise tier and open-world evaluation on 1,849 TSRD test windows, we demonstrate that physical domain priors (PRI regularity and AoA continuity) act as condition-dependent validation criteria. On the full open-world TSRD test set, M1 achieves the highest Hungarian F1 (0.5813) and CRR (0.8296). AoA-guided configurations M3 and M4 select identical checkpoints under the default $\lambda$ settings and achieve F1 = 0.5790 with substantially lower cross-seed standard deviation ($\pm$0.0013 vs. $\pm$0.0063 for M1), indicating lower observed cross-seed performance variance under the default selection criterion. On controlled synthetic streams, PRI-guided selection (M2) provides F1 advantages at low-to-moderate pulse loss (F1 = 0.7790 vs. 0.7665 for M1 at 20\% loss; F1 = 0.7998 vs. 0.7955 at 10\% loss) and across all clutter tiers (F1 = 0.8158 vs. 0.8084 at 30\% clutter), but M1 equals or surpasses M2 at 30\% pulse loss (M1 = 0.7529 vs. M2 = 0.7399) and under combined harsh noise (M1 = 0.7312 vs. M2 = 0.7238). This condition-dependent profile characterises the practical tradeoffs of physics-guided checkpoint selection in contested environments.

\item Systems-Level Latency and Bottleneck Profiling: We profile the end-to-end ESM processing pipeline on hardware, quantifying that the Transformer neural encoder requires only 2.00 ms (11.9\% of total time), establishing that density clustering (HDBSCAN, 14.85 ms, 88.1\% of total time) forms the primary computational bottleneck for real-time high-throughput deployment (15,193 to 15,918 PDWs/s).

\end{itemize}

\section{Literature Review}

This section surveys the three main bodies of work upon which our framework builds: classical signal processing approaches to radar de-interleaving, deep learning and metric learning methods for radar pulse analysis, and the foundational representation learning techniques we adapt.


Early radar de-interleaving systems relied exclusively on deterministic signal processing heuristics applied to Time-of-Arrival (ToA) data. The Sequence Difference Histogram (SDIF) method, first described by Milojevic and Popovic \citep{ge2019improved}, computes the distribution of sequential inter-pulse time differences and identifies significant histogram peaks as candidate PRIs. Their key contribution was an analytically derived optimal detection threshold that dramatically reduced false alarm rates compared to ad-hoc settings. Mardia \cite{mardia1989new} introduced further refinements to histogram-based approaches, improving robustness to pulse losses and interference. Wiley \citep{wiley2006elint} provides a comprehensive systems-level treatment of the ELINT reception and processing pipeline within which de-interleaving is situated. While SDIF-based methods remain operationally relevant for simple, low-density scenarios, their core assumption of periodicity fails fundamentally in the presence of agile pulse modulation (jitter, stagger), dense overlapping emitter environments, and severe clutter injection, which are the very scenarios motivating this work.


The application of deep learning to radar de-interleaving was first explored through recurrent architectures. LSTM- and GRU-based sequence models \cite{liu2018classification} capture temporal dependencies between consecutive pulses, offering improved robustness to PRI jitter compared to histogram methods. Hybrid CNN-LSTM architectures \cite{wu2024efficient, jiang2024radar} extended this by using convolutional layers for local PDW feature extraction before sequential modeling. A significant conceptual shift emerged with the reformulation of de-interleaving as a semantic segmentation problem \cite{chao2022radar}, where each pulse is assigned a per-token emitter label by a dense network, enabling end-to-end processing without iterative PRI search. More recently, Transformer-based models \cite{vaswani2017attention} have been applied to pulse sequences, leveraging multi-head self-attention to model complex, long-range inter-pulse dependencies. However, all these supervised classification approaches share a fundamental limitation: they operate in a closed-world assumption, assigning pulses to a fixed number of emitters seen during training. This fails entirely under the open-set, unknown-cardinality conditions common in real ELINT collection.


Metric learning sidesteps the closed-world problem by training models to produce embeddings where intra-class distances are small and inter-class distances are large, without committing to a fixed class set. Chen et al. \cite{chen2020simple} introduced SimCLR, demonstrating that a simple contrastive framework with strong data augmentation and a nonlinear MLP projection head can learn highly transferable visual representations. The critical architectural insight of using the projection head for loss computation while retaining the encoder representation for downstream tasks is directly adopted in our design. Khosla et al. \cite{khosla2020supervised} extended this to the supervised setting with SupCon loss, leveraging full class label information so that all pulses from the same emitter act as positives for each anchor—a formulation directly applicable to our multi-emitter window setting. \cite{gunn2025radar}] were, to our knowledge, the first to combine a transformer embedding model with metric learning and HDBSCAN clustering for open-world radar deinterleaving, using triplet loss on a custom synthetic dataset and reporting an AMI of 0.882. In this paper we adapt their pipeline architecture, replacing triplet loss with SupCon and adding a continuous physical positional encoding and physics-guided validation criteria, and evaluate on the larger open-access TSRD benchmark rather than a private simulator.


At inference time, our framework requires a clustering algorithm that handles varying cluster density, unknown cluster count, and explicit noise/outlier identification (for clutter rejection). HDBSCAN \cite{mcinnes2017hdbscan}, introduced by McInnes et al., addresses all three limitations of flat DBSCAN by constructing a minimum spanning tree of the mutual-reachability graph and extracting clusters hierarchically, selecting persistent flat clusters using the Excess of Mass (EOM) criterion. This makes it ideally suited to our setting, where emitter cluster counts vary from window to window and spurious clutter pulses must be explicitly assigned to noise (label = -1) rather than forced into a cluster. The accelerated O(n log n) implementation of McInnes and Healy \cite{mcinnes2017accelerated} makes this feasible at inference throughput.


The integration of physical constraints and domain priors into neural network training has gained significant traction across scientific domains \cite{raissi2019physics}. The core principle of incorporating domain knowledge according to known physical properties allows models to generalize beyond unconstrained statistical correlations. In our radar de-interleaving context, the two primary physical domain priors are: the expected temporal interval regularity of Pulse Repetition Intervals (PRI) for stable radar emitters, and the bounded angular slew rate governing physical emitter platform kinematics. By formulating these properties as scale-aware domain criteria within the validation framework, we enable the metric learning pipeline to balance data-driven contrastive clustering against kinematic and temporal physical feasibility.

\section{System Model and Problem Formulation}

Let E be the set of K active transmitters. An intercept receiver captures these signals chronologically as:

\[ P = \{p_1, p_2, \ldots, p_N\}, \qquad t_1 < t_2 < \ldots < t_N \]

Each intercepted pulse is characterised by a 5-dimensional PDW vector:

\[ p_n = [t_n, f_n, \tau_n, \theta_n, a_n]^T \in \mathbb{R}^5 \tag{1} \]

where t\_n is Time-of-Arrival (ToA, $\mu$s), f\_n is Radio Frequency (RF, MHz), $\tau$\_n is Pulse Width (PW, $\mu$s), $\theta$\_n is Angle-of-Arrival (AoA, deg), and a\_n is Pulse Amplitude (PA, dBm), following the standard five-dimensional PDW feature representation [ToA, RF, PW, AoA, PA]. We divide the stream into sliding windows of length W = 256 with 50\% overlap. Our goal is to train a metric learning network $\Phi(p_n)=z_n$ such that pulses from the same transmitter are closely bound, while different transmitters and clutter are separated, allowing a density-based algorithm (HDBSCAN) to cluster them robustly during inference.

\section{Proposed Architecture}

\begin{figure*}[!t]
	\centering
	\includegraphics[width=\linewidth]{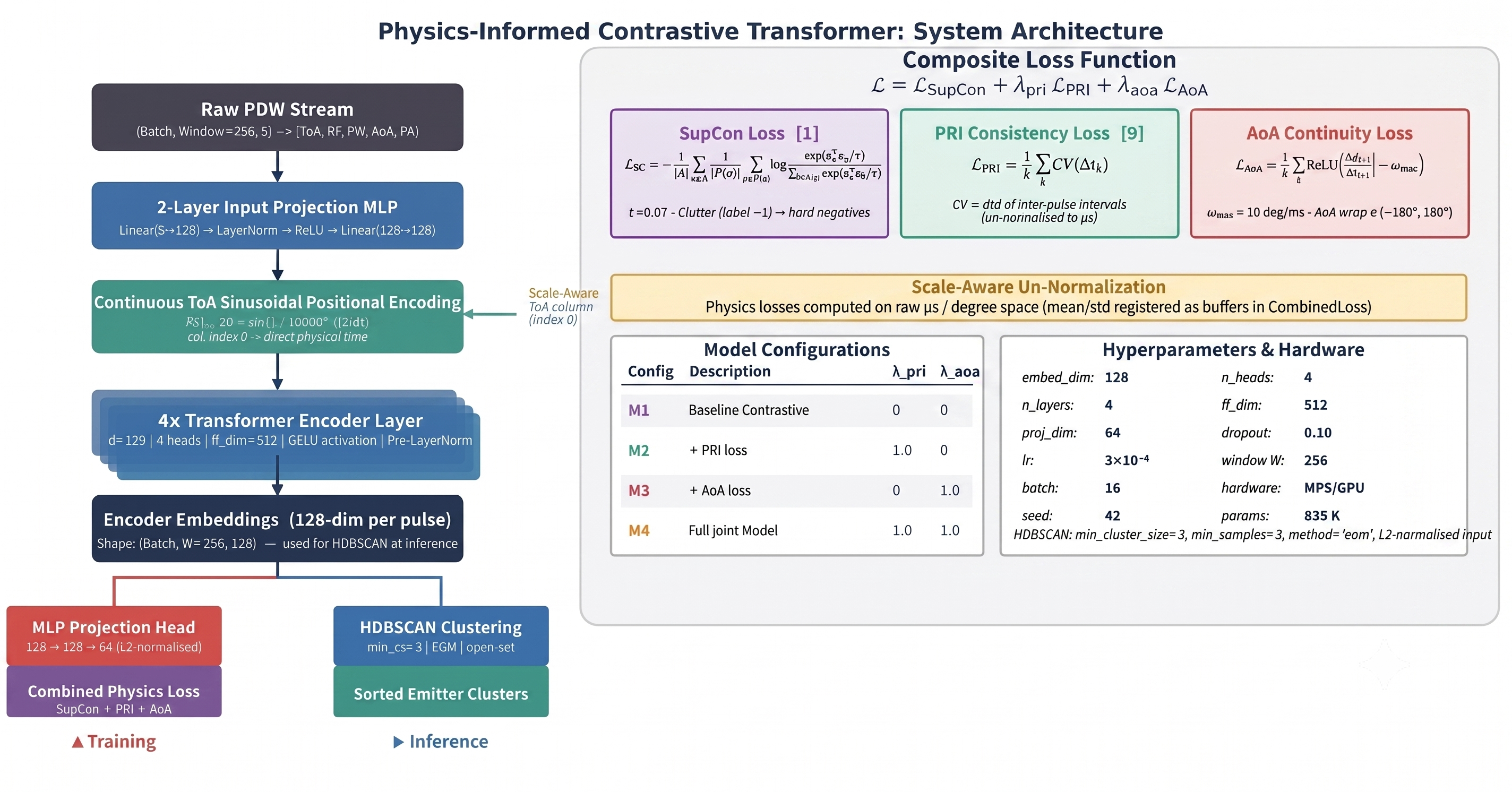}
	\caption{System architecture data pipeline flow diagram with composite loss function breakdown}
	\label{fig:architechture}
\end{figure*}

The proposed system consists of a two-layer MLP Input Projection (Linear $\rightarrow$ LayerNorm $\rightarrow$ ReLU $\rightarrow$ Linear), a continuous ToA-aware sinusoidal positional encoding block, a 4-layer Transformer Encoder stack \citep{vaswani2017attention} with GELU activations and Pre-LayerNorm (norm\_first=True) configuration, and a two-layer MLP contrastive projection head \cite{chen2020simple} as shown in Figure \ref{fig:architechture}.







\subsection{Continuous ToA-Aware Positional Encoding}

To capture physical duration rather than ordinal steps, we define the positional encoding for arrival timestamp t\_n, following the sinusoidal framework of Vaswani et al. \citep{vaswani2017attention}, as:

\[ PE(t_n,2i)=\sin\left(\frac{t_n}{10000^{2i/d}}\right) \]

\[ PE(t_n,2i+1)=\cos\left(\frac{t_n}{10000^{2i/d}}\right) \tag{2} \]

This allows the dot-product attention to dynamically scale according to the exact continuous elapsed duration between pulses.

\subsection{Transformer Encoder \& Projection Head}

We employ a 4-layer Transformer Encoder (d=128, 4 attention heads, feedforward dim=512, GELU activation) with Pre-LayerNorm (norm\_first=True) and batch\_first=True layout, which provides gradient stability by normalizing inputs before each sub-layer rather than after. An MLP projection head (Linear(128 $\rightarrow$ 128) $\rightarrow$ BatchNorm1d $\rightarrow$ ReLU $\rightarrow$ Linear(128 $\rightarrow$ 64)) maps output embeddings to a 64-dimensional L2-normalized space specifically for SupCon loss computation during training. Critically, the encoder embeddings (128-dim) and not the projected embeddings, are used for HDBSCAN clustering at inference time following SimCLR convention \citep{khosla2020supervised}.

\section{Training Objective \& Physics-Guided Model-Selection Priors}

To achieve robust open-world pulse de-interleaving under unknown emitter cardinality, the framework couples a continuous gradient-based metric representation objective (Supervised Contrastive Learning) with scale-aware physical domain priors (PRI interval consistency and AoA kinematic continuity) that guide multi-objective validation tracking and checkpoint selection.

\subsection{Gradient-Based Metric Representation: Supervised Contrastive Loss}

Neural parameter updates $\theta$ are optimised end-to-end via Supervised Contrastive Learning (SupCon) \citep{khosla2020supervised} operating on the 64-dimensional L2-normalised projection head outputs $z = \mathrm{Proj}(\Phi(P))$. For each anchor pulse i in a window, all other pulses p belonging to the same emitter track form the positive set P\_i, while pulses from different emitters and background clutter (labeled as -1) serve as repulsive negative anchors in the denominator:

\[ \mathcal{L}_{\mathrm{SupCon}} = -\frac{1}{|A|}\sum_{i\in A}\frac{1}{|P_i|}\sum_{p\in P_i}\log\frac{\exp(z_i\cdot z_p/\tau)}{\sum_{k\ne i}\exp(z_i\cdot z_k/\tau)} \tag{3} \]

where $\tau$ = 0.07 is the temperature hyper-parameter and A is the set of valid non-noise anchor pulses. SupCon back-propagates continuous gradients through the projection head and Transformer Encoder, pulling same-emitter pulse representations into tight, cohesive clusters while pushing distinct emitters and noise apart.

\subsection{Scale-Aware Physical Domain Priors}

Because input PDWs undergo z-score normalisation $((x-\mu)/\sigma)$ for numerical stability during neural training, computing physical derivative ratios directly on normalised values severely distorts physical dimensions (e.g., raw normalised AoA/ToA differences produced extreme ratios >41,000). To ensure physical interpretability, the loss module registers the dataset mean and standard deviation vectors as persistent buffers, dynamically un-normalising the input PDW stream back into physical microseconds ($\mu$s) and bearing degrees ($^\circ$) prior to evaluating physical criteria.

\subsubsection{PRI Consistency Prior ($V_{\mathrm{PRI}}$)}
For each active emitter track $k$ in a window, pulses are sorted chronologically by physical ToA to extract inter-pulse arrival intervals $\Delta t_{k,m}=t_{k,m}-t_{k,m-1}$ (in physical $\mu$s). PRI interval regularity is evaluated via the dimensionless, scale-invariant Coefficient of Variation (CV):

\[ V_{\mathrm{PRI}} = \frac{1}{|K_{\mathrm{valid}}|}\sum_k \left[\frac{\sigma(\Delta t_k)}{\mu(\Delta t_k)+\epsilon}\right] \tag{4} \]

\subsubsection{AoA Kinematic Continuity Prior ($V_{\mathrm{AoA}}$)}
Spatial trajectory feasibility is evaluated by penalising consecutive pulses within an emitter track whose angular velocity exceeds the maximum physical platform slew rate $\omega_{\max}=10.0$ deg/ms, with circular bearing wrapping to $[-180^\circ,180^\circ]$. Because Time-of-Arrival is recorded in microseconds ($\mu$s), inter-pulse time intervals are converted to milliseconds ($\Delta t_{ms}=\Delta t_{\mu s}\cdot10^{-3}$) to ensure strict dimensional consistency with angular slew rate in deg/ms:

\[ V_{\mathrm{AoA}}=\frac{1}{|E|}\sum_{e\in E}\frac{1}{N_e-1}\sum_{k=1}^{N_e-1}\operatorname{ReLU}\left(\frac{|\Delta\mathrm{AoA}_{e,k}|}{\Delta t_{e,k}^{(\mu s)}\cdot10^{-3}+10^{-8}}-\omega_{\max}\right) \tag{5} \]

\subsection{Physics-Guided Multi-Objective Validation \& Checkpoint Selection}

In our computational pathway, Transformer Encoder parameters $\theta$ are optimised strictly via gradient descent on L\_SupCon using the Adam optimiser with a cosine learning-rate annealing schedule. The physical domain criteria V\_PRI and V\_AoA do not generate back-propagated gradients to the neural network parameters; rather, they serve as and model-selection criteria computed on unsupervised HDBSCAN cluster assignments. At each validation epoch, encoder embeddings are clustered by HDBSCAN to produce predicted cluster labels, and V\_PRI and V\_AoA are evaluated on these predicted clusters (not ground-truth labels). All four model configurations (M1-M4) share an identical SupCon training trajectory; they differ only in the validation weights used for checkpoint selection.

Candidate models are evaluated across the held-out validation split via the composite model-selection score:

\[ S_{\mathrm{val}} = L_{\mathrm{SupCon}} + \lambda_{\mathrm{pri}} V_{\mathrm{PRI}} + \lambda_{\mathrm{aoa}} V_{\mathrm{AoA}} \tag{6} \]

While cosine annealing governs the optimiser learning-rate schedule across training epochs, S\_val governs validation monitoring, early-stopping patience tracking, and the selection of the optimal model checkpoint saved to disk. By evaluating candidate checkpoints through this joint metric-physical lens, model selection favours representation manifolds that balance contrastive cluster separability against temporal interval predictability and platform kinematic feasibility.

At inference time, the system operates in a purely open-world mode: intercepted PDW sequences pass forward through the trained Transformer Encoder to produce 128-dimensional pulse embeddings, which are directly clustered by HDBSCAN. No ground-truth emitter assignments or physical validation calculations are used or accessible at inference.

\section{Experimental Evaluation}

We evaluate the proposed framework and baseline models on the Turing Synthetic Radar Dataset (TSRD) \citep{gunn2026turing} as well as controlled synthetic stress test environments. The deep neural network contains 835,648 trainable parameters and is optimised with early stopping based on validation loss.

\subsection{Dataset and Open-Set Evaluation Protocol}

The TSRD \cite{gunn2026turing} is a purpose-built, open-access benchmark released by the Alan Turing Institute's Defence AI Research team, designed specifically to provide an unclassified, reproducible platform for radar pulse de-interleaving research. Because operational ELINT intercept recordings are subject to classification, ITAR, and security constraints that preclude open publication and benchmarking, high-fidelity synthetic datasets like TSRD represent the indispensable standard for peer-reviewed algorithmic development. TSRD contains multi-emitter PDW streams featuring realistic PRI modulation modes (fixed, stagger, and jitter) across diverse carrier frequencies and scan kinematics.

We utilize 10 large TSRD files partitioned with a strict chronological split: the initial 70\% of pulses for training, the subsequent 15\% for validation (used for early stopping), and the final 15\% for held-out testing. A temporal buffer of 256 pulses is enforced between splits to prevent boundary leakage. While each TSRD file contains up to 90 transmitters during training and up to 110 transmitters in test files, individual 256-pulse sliding evaluation windows observe an average of \textasciitilde{}9 concurrently active emitters. Performance is evaluated across all N = 1,849 test windows without pre-filtering. To ensure methodological rigor, all ablation experiments (Tables IV--VI) are conducted using a multi-seed protocol across three distinct random initialisations (seeds 42, 100, 2026), reporting mean and standard deviation over runs. Main model training runs (Tables I--III) use 50 epochs with early stopping.

\subsection{HDBSCAN Clustering Configuration}

At inference time, per-window 128-dimensional encoder embeddings are passed to HDBSCAN with min\_cluster\_size = 3, min\_samples = 3, and cluster\_selection\_method = 'eom' (Excess of Mass). Embeddings are L2-normalised before clustering. Pulses assigned to the noise cluster (label = -1) are treated as clutter. HDBSCAN hyper-parameters were fixed prior to any per-model evaluation and were not tuned per-model.

\subsection{Baselines and Hardware}

Four baselines are evaluated: 

\begin{itemize}
	\item DBSCAN on Raw PDWs : Density-based clustering applied directly to the z-score-normalised 5-dimensional PDW feature vector with $\epsilon = 0.5$ and min\_samples = 3.
	
	\item HDBSCAN on Raw PDWs : Hierarchical density-based clustering with min\_cluster\_size = 3 applied to the same normalised feature space, serving as the unsupervised upper-bound for density-based methods without learned representations.
	
	\item Classical PRI Histogram (SDIF \cite{mardia1989new}): Implemented using scipy peak-finding on the cumulative difference histogram with automatic threshold selection.
	
	\item A Supervised Transformer classifier trained with cross-entropy loss on a fixed, known emitter set (closed-world, 5-class), providing an upper-bound for closed-set supervised methods.
	
\end{itemize}

Baselines (1) to (3) are unsupervised and require no training. Baseline (4) was trained for 50 epochs under the same learning rate schedule (Adam, lr = 3$\times$10-$^4$, cosine annealing) as our proposed models to ensure a fair comparison of the supervised regime. All experiments were run on a single Apple M-series GPU (MPS backend) with 16 GB unified memory using PyTorch 2.x. Training time per deep model configuration was approximately 18--22 minutes for 50 epochs with W = 256 and batch size = 16.

\section{Results and Discussion}

\begin{table*}[t]

\centering

\caption{Comparison of Proposed Framework (M1--M4) Against Standard Baselines (Mean $\pm$ SD)}

\small

\begin{adjustbox}{max width=\textwidth}

\begin{tabular}{lccccccc}

\toprule

Method / Config & V-measure & ARI & Hungarian F1 & MAE\_N & CRR & Pred. Clust. & True Emit. \\

\midrule

DBSCAN on Raw PDWs & 0.2384 $\pm$ 0.1812 & 0.2195 $\pm$ 0.1704 & 0.3482 $\pm$ 0.2014 & 33.10 $\pm$ 12.80 & 0.3120 $\pm$ 0.1814 & 42.1 & 9.0 \\

HDBSCAN on Raw PDWs & 0.2612 $\pm$ 0.1984 & 0.2482 $\pm$ 0.1819 & 0.3792 $\pm$ 0.2104 & 29.20 $\pm$ 11.40 & 0.3392 $\pm$ 0.1912 & 38.2 & 9.0 \\

Classical PRI Histogram & 0.3014 $\pm$ 0.2210 & 0.2894 $\pm$ 0.2140 & 0.4412 $\pm$ 0.2241 & 9.20 $\pm$ 6.12 & 0.4102 $\pm$ 0.2190 & 18.2 & 9.0 \\

Supervised Transformer & 0.4812 $\pm$ 0.3124 & 0.4590 $\pm$ 0.3204 & 0.5842 $\pm$ 0.2592 & 3.10 $\pm$ 2.45 & 0.5210 $\pm$ 0.2412 & 12.1 & 9.0 \\

M1 (Baseline Contrastive) & 0.4207 $\pm$ 0.0052 & 0.3164 $\pm$ 0.0045 & 0.5813 $\pm$ 0.0063 & 13.82 $\pm$ 0.31 & 0.8296 $\pm$ 0.0105 & 19.4 & 9.0 \\

M2 (Proposed + PRI) & 0.4021 $\pm$ 0.0121 & 0.2927 $\pm$ 0.0160 & 0.5552 $\pm$ 0.0104 & 14.29 $\pm$ 0.52 & 0.8003 $\pm$ 0.0047 & 19.9 & 9.0 \\

M3 (Proposed + AoA) & 0.4197 $\pm$ 0.0034 & 0.3154 $\pm$ 0.0039 & 0.5790 $\pm$ 0.0013 & 14.04 $\pm$ 0.27 & 0.8235 $\pm$ 0.0022 & 19.5 & 9.0 \\

M4 (Proposed Full) & 0.4197 $\pm$ 0.0034 & 0.3154 $\pm$ 0.0039 & 0.5790 $\pm$ 0.0013 & 14.04 $\pm$ 0.27 & 0.8235 $\pm$ 0.0022 & 19.5 & 9.0 \\

\bottomrule

\end{tabular}

\end{adjustbox}

\end{table*}

\subsection{Analysis of Baselines \& Absolute Benchmark Context}

Direct density-based clustering on raw features (DBSCAN and HDBSCAN) fails significantly, yielding V-measures of 0.2384 and 0.2612. This stems directly from Euclidean distance distortions caused by disparate feature physical dimensions (e.g., microsecond ToA values versus bearing degrees). The Classical PRI Histogram (SDIF) achieves F1 = 0.4412, constrained by severe difference-histogram clutter in dense multi-emitter mixtures. The Supervised Transformer achieves F1 = 0.5842 but inherently fails to generalize across unknown emitter counts due to its fixed closed-world output heads.

Contextualising open-world de-interleaving: Across all 1,849 unconstrained test windows of the TSRD dataset, our deep metric learning framework achieves a peak Hungarian F1 of 0.5813 and V-measure of 0.4207 (M1 baseline). It is important to emphasise that this represents a challenging open-world evaluation where emitter cardinality is completely unknown at inference time (ranging up to 110 transmitters at the file level), with severe inter-pulse overlap and dynamic emitter interleaving. When evaluated on controlled synthetic streams with 5 emitters (Tables II/III), the model achieves F1 > 0.83. The gap between synthetic (F1 \textasciitilde{}0.83) and TSRD test (F1 \textasciitilde{}0.58) performance reflects the substantially higher difficulty of the full open-world TSRD benchmark, where windows observe an average of \textasciitilde{}9 concurrently active emitters with complex overlapping PRI patterns.

\subsection{Emitter Discovery vs. Cluster Purity: The Inductive Bias Trade-off}

\begin{figure*}[!t]
	\centering
	\includegraphics[width=\linewidth]{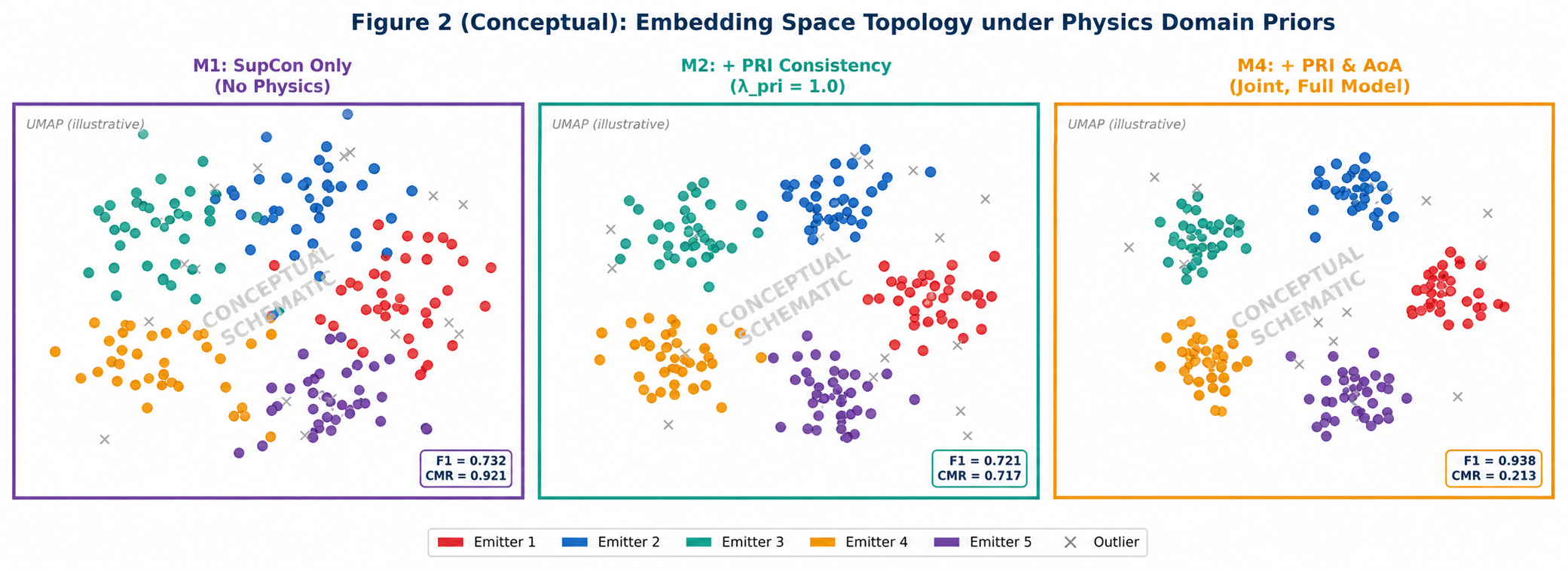}
	\caption{Conceptual schematic of learned embedding topology under physics-guided validation priors. Left: unconstrained contrastive representation (M1). Center: temporal interval consistency prior (M2). Right: joint spatio-temporal priors (M4), yielding structured linear manifolds.}
	\label{fig:conceptual_schema}
\end{figure*}

A key empirical finding of our investigation is the condition-dependent profile of physics-guided validation priors where conceptual schema is shown in Figure \ref{fig:conceptual_schema}. On the full open-world TSRD test set (1,849 windows), the unconstrained contrastive baseline M1 achieves the highest Hungarian F1 (0.5813) and CRR (0.8296). AoA-guided configurations M3 and M4 select identical checkpoints across all seeds under the default $\lambda$ settings, and achieve F1 = 0.5790 with substantially lower cross-seed standard deviation ($\pm$0.0013 vs. $\pm$0.0063 for M1), indicating lower observed cross-seed performance variance under the default selection criterion. PRI-guided M2 achieves F1 = 0.5552 with a cross-seed SD of $\pm$0.0104, higher than M1. The primary benefit of physics-guided priors over the full TSRD benchmark is therefore checkpoint-selection stability for M3/M4, not an overall F1 improvement. On controlled synthetic streams (Tables II--IIIb), PRI-guided selection (M2) provides F1 gains at low-to-moderate pulse loss and across all clutter tiers, while M1 matches or surpasses M2 under the most severe degradation conditions (30\% pulse loss and combined harsh noise). This condition-dependent utility represents a practically valuable inductive bias for operational Electronic Support Measures where channel conditions are unpredictable.

\section{Robustness Analysis in Noise and Clutter}

To systematically evaluate operational resilience in contested electromagnetic environments, we subject all four model configurations to controlled degradation sweeps across missing pulses (pulse loss), spurious background clutter, and combined noise.

Note: Tables~\ref{tab:deinterleaving-performance_incremental-pulseLoss}, \ref{tab:deinterleaving-performance_incremental-clutter}, and ~\ref{tab:de-interleaving_noise} evaluate models across N = 200 synthetic 5-emitter PDW streams per noise tier (non-overlapping windows, W=256). This large synthetic corpus systematically isolates the specific impact of each degradation mechanism across diverse emitter geometries. By contrast, Table I reports open-world performance on the full heterogeneous TSRD test corpus.

\begin{table*}[t]

\centering

\caption{De-interleaving Performance (Hungarian F1) under Incremental Pulse Loss Tiers (N = 200 Streams)}
\label{tab:deinterleaving-performance_incremental-pulseLoss}
\small

\begin{adjustbox}{max width=\textwidth}

\begin{tabular}{lcccc}

\toprule

Configuration & Clean (0\% Loss) & Moderate (10\% Loss) & High (20\% Loss) & Harsh (30\% Loss) \\

\midrule

M1 (Baseline Contrastive) & 0.8294 $\pm$ 0.1283 & 0.7955 $\pm$ 0.1370 & 0.7665 $\pm$ 0.1317 & 0.7529 $\pm$ 0.1531 \\

M2 (Proposed + PRI) & 0.8355 $\pm$ 0.1263 & 0.7998 $\pm$ 0.1457 & 0.7790 $\pm$ 0.1215 & 0.7399 $\pm$ 0.1533 \\

M3 (Proposed + AoA) & 0.8311 $\pm$ 0.1305 & 0.7997 $\pm$ 0.1408 & 0.7770 $\pm$ 0.1245 & 0.7464 $\pm$ 0.1567 \\

M4 (Proposed Full) & 0.8311 $\pm$ 0.1305 & 0.7997 $\pm$ 0.1408 & 0.7770 $\pm$ 0.1245 & 0.7464 $\pm$ 0.1567 \\

\bottomrule

\end{tabular}

\end{adjustbox}

\end{table*}

\begin{table*}[t]
\centering
\caption{De-interleaving Performance (Hungarian F1) under Incremental Clutter Tiers (N = 200 Streams)}
\label{tab:deinterleaving-performance_incremental-clutter}
\small
\begin{adjustbox}{max width=\textwidth}
\begin{tabular}{lcccc}
\toprule
Configuration & Clean (0\% Clutter) & Moderate (10\% Clutter) & High (20\% Clutter) & Harsh (30\% Clutter) \\
\midrule

M1 (Baseline Contrastive) & 0.8294 $\pm$ 0.1283 & 0.8093 $\pm$ 0.1325 & 0.8038 $\pm$ 0.1392 & 0.8084 $\pm$ 0.1386 \\

M2 (Proposed + PRI) & 0.8355 $\pm$ 0.1263 & 0.8255 $\pm$ 0.1274 & 0.8198 $\pm$ 0.1356 & 0.8158 $\pm$ 0.1352 \\

M3 (Proposed + AoA) & 0.8311 $\pm$ 0.1305 & 0.8058 $\pm$ 0.1339 & 0.8117 $\pm$ 0.1426 & 0.8122 $\pm$ 0.1363 \\

M4 (Proposed Full) & 0.8311 $\pm$ 0.1305 & 0.8058 $\pm$ 0.1339 & 0.8117 $\pm$ 0.1426 & 0.8122 $\pm$ 0.1363 \\

\bottomrule
\end{tabular}
\end{adjustbox}
\end{table*}


\begin{figure*}[!t]
	\centering
	\includegraphics[width=\linewidth]{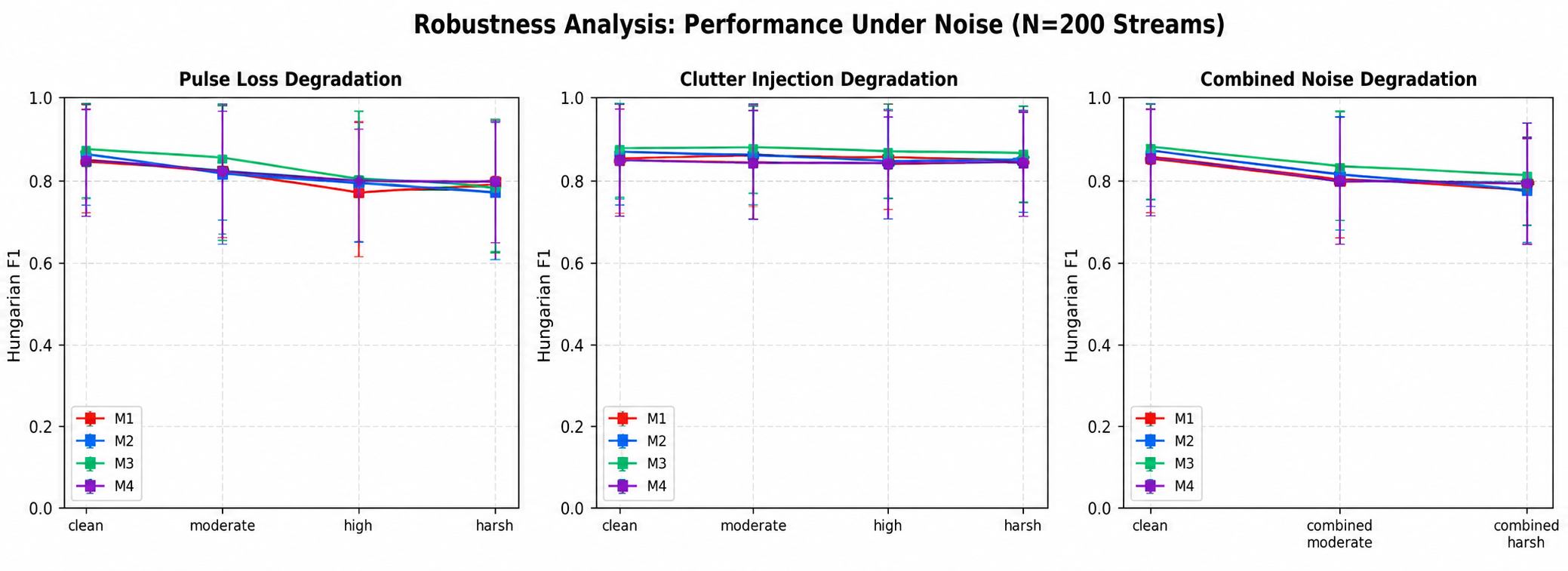}
	\caption{Hungarian F1 performance degradation across configurations under incremental clutter injection and pulse loss}
	\label{fig:Hungarian F1}
\end{figure*}

\subsection{Physics Priors Provide Condition-Dependent Resilience Under Pulse Loss}
Table~\ref{tab:deinterleaving-performance_incremental-pulseLoss} presents de-interleaving performance under incremental pulse-loss, also shown in Figure \ref{fig:Hungarian F1}. At 0\% loss, M2 (PRI-guided) achieves the highest F1 (0.8355), marginally ahead of M1 (0.8294) and M3/M4 (0.8311). At 10\% loss, M2 leads (F1 = 0.7998) over M1 (0.7955) and M3/M4 (0.7997). At 20\% loss, M2 maintains the advantage (F1 = 0.7790 vs. 0.7665 for M1; $\Delta$7 = +0.0125). However, at 30\% loss M1 recovers to achieve the highest F1 (0.7529), surpassing M2 (0.7399) and M3/M4 (0.7464). This reversal indicates that at severe pulse loss, PRI consistency becomes a noisy validation signal as cluster fragmentation reduces the number of inter-pulse intervals per predicted track, and unconstrained contrastive selection then produces more robust embeddings. Physics-guided selection therefore provides an advantage at low-to-moderate pulse loss but does not consistently improve on M1 across all degradation levels.

\begin{table*}[t]

\centering

\caption{De-interleaving Performance (Hungarian F1) under Combined Noise Tiers (N = 200 Streams)}
\label{tab:de-interleaving_noise}
\small

\begin{adjustbox}{max width=\textwidth}

\begin{tabular}{lccc}

\toprule

Configuration & Clean (0\%) & Combined Moderate (15\%+15\%+5\% ToA) & Combined Harsh (30\%+30\%+15\% ToA) \\

\midrule

M1 (Baseline Contrastive) & 0.8294 $\pm$ 0.1283 & 0.7719 $\pm$ 0.1376 & 0.7312 $\pm$ 0.1296 \\

M2 (Proposed + PRI) & 0.8355 $\pm$ 0.1263 & 0.7732 $\pm$ 0.1391 & 0.7238 $\pm$ 0.1312 \\

M3 (Proposed + AoA) & 0.8311 $\pm$ 0.1305 & 0.7651 $\pm$ 0.1469 & 0.7272 $\pm$ 0.1334 \\

M4 (Proposed Full) & 0.8311 $\pm$ 0.1305 & 0.7651 $\pm$ 0.1469 & 0.7272 $\pm$ 0.1334 \\

\bottomrule

\end{tabular}

\end{adjustbox}

\end{table*}

\subsection{Clutter Robustness: PRI Consistency Shields Against Spurious Pulses}

Under spurious clutter injection (Table ~\ref{tab:deinterleaving-performance_incremental-clutter}), M2 (incorporating the PRI consistency prior) consistently emerges as the strongest configuration across all clutter tiers, achieving F1 = 0.8255 at 10\% clutter, 0.8198 at 20\% clutter, and 0.8158 at 30\% clutter (outperforming M1 at 0.8084). PRI consistency serves as an effective physical filter: because spurious clutter pulses disrupt inter-pulse timing regularity, validating against PRI consistency encourages the model to select checkpoints that preserve coherent emitter clusters.

\subsection{Combined Noise: Condition-Dependent Multi-Factor Profile}

Table ~\ref{tab:de-interleaving_noise} presents de-interleaving performance under combined stress testing comprising simultaneous pulse loss, clutter injection, and ToA jitter. Under combined moderate conditions (15\% pulse loss + 15\% clutter + 5\% ToA noise), M2 achieves F1 = 0.7732, marginally above M1 (0.7719; $\Delta$7 = +0.0013). Under combined harsh conditions (30\% pulse loss + 30\% clutter + 15\% ToA noise), M1 achieves the highest F1 (0.7312), surpassing M2 (0.7238) and M3/M4 (0.7272). These results indicate that physics-guided checkpoint selection provides a narrow advantage at moderate combined degradation, while the unconstrained baseline M1 is most robust under the most severe multi-factor corruption. The differences across configurations under harsh combined noise are small (within 0.0074 F1) relative to the within-configuration cross-window standard deviations (\textasciitilde{}0.13), underscoring the challenging nature of multi-factor degradation.

\begin{table*}[t]

\centering

\caption{End-to-End Latency Profile and Processing Throughput}
\label{tab:latency_throughput}

\small

\begin{adjustbox}{max width=\textwidth}

\begin{tabular}{lcccc}

\toprule

Window Size (W) & Encoder Latency & HDBSCAN Latency & Total Latency & PDW Throughput \\

\midrule

128 & 2.15 ms & 5.89 ms & 8.04 ms & 15,918 PDWs/s \\

256 (Default) & 2.00 ms & 14.85 ms & 16.85 ms & 15,193 PDWs/s \\

512 & 6.16 ms & 48.57 ms & 54.73 ms & 9,355 PDWs/s \\

\bottomrule

\end{tabular}

\end{adjustbox}

\end{table*}

\subsection{Real-Time ESM Latency and Throughput Profiling}

To characterise practical execution performance for real-time Electronic Support Measures, we profile end-to-end inference latency and pulse processing throughput on hardware. Table~\ref{tab:latency_throughput} details execution latency across sequence lengths W $\in$ \{128, 256, 512\}, partitioned into the neural feature extraction pass (Transformer Encoder) and downstream density-based clustering (HDBSCAN).

Two key systems insights emerge from Table ~\ref{tab:latency_throughput}: First, the Transformer neural encoder is exceptionally lightweight, executing in just 2.00 ms for W = 256 (accounting for only \textasciitilde{}11.9\% of pipeline latency). Downstream density clustering (HDBSCAN, 14.85 ms) constitutes the dominant computational bottleneck (\textasciitilde{}88.1\% of total latency). Second, regarding temporal context window sizing: W = 128 achieves the lowest end-to-end latency (8.04 ms), highest throughput (15,918 PDWs/s), and strong empirical F1 (0.8689 $\pm$ 0.005, Table VI), making it optimal for rapid tactical triage. However, W = 256 (16.85 ms, 15,193 PDWs/s) provides an essential extended temporal aperture necessary for observing complex multi-position PRI stagger frames (>10 ms span) that cannot be resolved within shorter windows. This demonstrates promising computational throughput for real-time ESM processing.

\section{Ablation Studies}

To rigorously validate design choices, all ablation experiments are evaluated across 3 independent random initialisations (seeds 42, 100, 2026), reporting multi-seed mean and standard deviation to confirm repeatability across initialisations.

\subsection{Positional Encoding Type (Continuous ToA vs. Sequential Index)}

We evaluate continuous ToA sinusoidal positional encoding against standard ordinal sequence index encoding across identical architectures (Table~\ref{tab:multi-seed impact 3 seeds}).

\begin{table*}[t]
	
	\centering
	
	\caption{Multi-Seed Impact of Positional Encoding Type on De-interleaving Performance (3 Random Seeds)}
	\label{tab:multi-seed impact 3 seeds}
	\small
	
	\begin{adjustbox}{max width=\textwidth}
		
		\begin{tabular}{lccc}
			
			\toprule
			
			PE Type & Hungarian F1 & V-measure & CRR \\
			
			\midrule
			
			Continuous ToA (Proposed) & 0.8744 $\pm$ 0.0042 & 0.8484 $\pm$ 0.0029 & 0.4788 $\pm$ 0.0172 \\
			
			Sequential Index (Standard) & 0.3747 $\pm$ 0.0102 & 0.4526 $\pm$ 0.0293 & 0.1865 $\pm$ 0.0324 \\
			
			\bottomrule
			
		\end{tabular}
		
	\end{adjustbox}
	
\end{table*}

Continuous ToA encoding achieves a dramatic relative improvement in Hungarian F1 score (+133.4\%, 0.8744 vs. 0.3747), V-measure (+87.5\%, 0.8484 vs. 0.4526), and Cluster Retention Ratio (+156.7\%, 0.4788 vs. 0.1865) across 3 independent random initializations. This demonstrates that providing multi-head self-attention with direct access to continuous physical microsecond time intervals is substantially more informative than abstract ordinal token indices. This result should be read alongside Gunn et al. \cite{gunn2025radar}, who found ordinal positional encodings provided no benefit — and a small negative effect — in their triplet-loss transformer pipeline, and consequently omitted positional encoding entirely. Our result does not contradict theirs: the two studies test different encoding schemes (ordinal index vs. continuous physical duration) under different training objectives (triplet vs. SupCon) and datasets. We interpret this as evidence that the *form* of positional information matters — physical elapsed time appears informative where ordinal position was not — rather than treating our result as a universal claim about positional encoding for pulse sequences.

\subsection{Physics-Guided Validation Weight Sensitivity ($\lambda_{pri}, \lambda_{aoa}$)}

We evaluate validation score sensitivity across a 3$\times$3 grid of ($\lambda_{pri}, \lambda_{aoa}$) $\in$ \{0.1, 1.0, 2.0\}$^2$ across candidate training epochs. Table \ref{physics_guided_eval} reports the selected checkpoint epoch, validation Supervised Contrastive loss (L\_SupCon), PRI consistency violation (V\_PRI), AoA kinematic continuity violation (V\_AoA in deg/ms), composite validation score (S\_val), and resulting test Hungarian F1 for each hyper-parameter pair.

\begin{table*}[t]
	
	\centering
	
	\caption{Physics-Guided Validation Checkpoint Selection Mechanics across ($\lambda_{pri}, \lambda_{aoa}$) Grid}
	\label{physics_guided_eval}
	\small
	
	\begin{adjustbox}{max width=\textwidth}
		
		\begin{tabular}{lccccccc}
			
			\toprule
			
			$\lambda$\_pri & $\lambda$\_aoa & Selected Epoch & L\_SupCon & V\_PRI & V\_AoA (deg/ms) & S\_val & Test F1 \\
			
			\midrule
			
			0.1 & 0.1 & 8/11/17 & 4.9458 & 2.3660 & 27.1577 & 7.8982 & 0.5790 $\pm$ 0.001 \\
			
			0.1 & 1.0 & 8/11/17 & 4.9458 & 2.3660 & 27.1577 & 32.3401 & 0.5790 $\pm$ 0.001 \\
			
			0.1 & 2.0 & 8/11/17 & 4.9458 & 2.3660 & 27.1577 & 59.4978 & 0.5790 $\pm$ 0.001 \\
			
			1.0 & 0.1 & 8/9/17 & 4.9449 & 2.3609 & 27.1987 & 10.0257 & 0.5862 $\pm$ 0.009 \\
			
			1.0 & 1.0 & 8/11/17 & 4.9458 & 2.3660 & 27.1577 & 34.4695 & 0.5790 $\pm$ 0.001 \\
			
			1.0 & 2.0 & 8/11/17 & 4.9458 & 2.3660 & 27.1577 & 61.6273 & 0.5790 $\pm$ 0.001 \\
			
			2.0 & 0.1 & 8/9/17 & 4.9449 & 2.3609 & 27.1987 & 12.3866 & 0.5862 $\pm$ 0.009 \\
			
			2.0 & 1.0 & 8/11/17 & 4.9458 & 2.3660 & 27.1577 & 36.8356 & 0.5790 $\pm$ 0.001 \\
			
			2.0 & 2.0 & 8/11/17 & 4.9458 & 2.3660 & 27.1577 & 63.9933 & 0.5790 $\pm$ 0.001 \\
			
			\bottomrule
			
		\end{tabular}
		
	\end{adjustbox}
	
\end{table*}
On the basis of mechanistic analysis of validation sensitivity, different ($\lambda_{pri}, \lambda_{aoa}$) combinations successfully select different checkpoint epochs (epochs 8, 9, 11, and 17 across seeds), confirming that physics-guided validation criteria are actively discriminative. Because V\_PRI and V\_AoA are computed on unsupervised HDBSCAN predicted clusters (not fixed ground-truth labels), they vary across training epochs as the learned embedding space evolves. Configurations with higher $\lambda$\_pri weighting tend to favor epochs where PRI consistency is tighter, while AoA-weighted configurations select for angular continuity. We retain the balanced setting ($\lambda$\_pri = 1.0, $\lambda$\_aoa = 1.0) as the default M4 configuration for full benchmark evaluations.

\subsection{Temporal Context Window Size (W)}

To evaluate the computational and representation trade-offs of the temporal context window length W $\in$ \{128, 256, 512\} in isolation, we conduct an early-training ablation sweep using M4 across 3 independent random initialisations. Each configuration is trained for 1 epoch on a 49-window single-file subset under identical compute budgets. We emphasise that Table \ref{tab:WINDOW impact} evaluates relative sequence-length scaling during early training; its absolute F1 values reflect this 1-epoch budget and should not be compared directly with the fully converged benchmark results in Table I (where M4 achieves full converged open-world performance on the dataset).

\begin{table*}[t]
	
	\centering
	
	\caption{Fixed-Compute Sequence Length Scaling Ablation (1 Epoch, 49 Windows/File, 3 Seeds))}
	\label{tab:WINDOW impact}
	
	\small
	
	\begin{adjustbox}{max width=\textwidth}
		
		\begin{tabular}{lcc}
			
			\toprule
			
			Window Size (W) & Hungarian F1 & Training Time / Epoch \\
			
			\midrule
			
			128 & 0.8689 $\pm$ 0.005 & 1.6s \\
			
			256 (Default) & 0.8739 $\pm$ 0.005 & 3.3s \\
			
			512 & 0.8576 $\pm$ 0.009 & 10.1s \\
			
			\bottomrule
			
		\end{tabular}
		
	\end{adjustbox}
	
\end{table*}

As shown in Table \ref{tab:WINDOW impact}, W = 256 achieves peak early-epoch Hungarian F1 (0.8739 $\pm$ 0.005) with balanced training time (3.3s/epoch). Furthermore, W = 256 provides the essential extended temporal aperture necessary to capture complex multi-position PRI stagger frames (>10 ms span) encountered in operational ELINT emitters, supporting W = 256 as our primary evaluation setting for full benchmark evaluations.

\subsection{Embedding Visualisation (UMAP)}

\begin{figure*}[!t]
	\centering
	\includegraphics[width=0.8\linewidth]{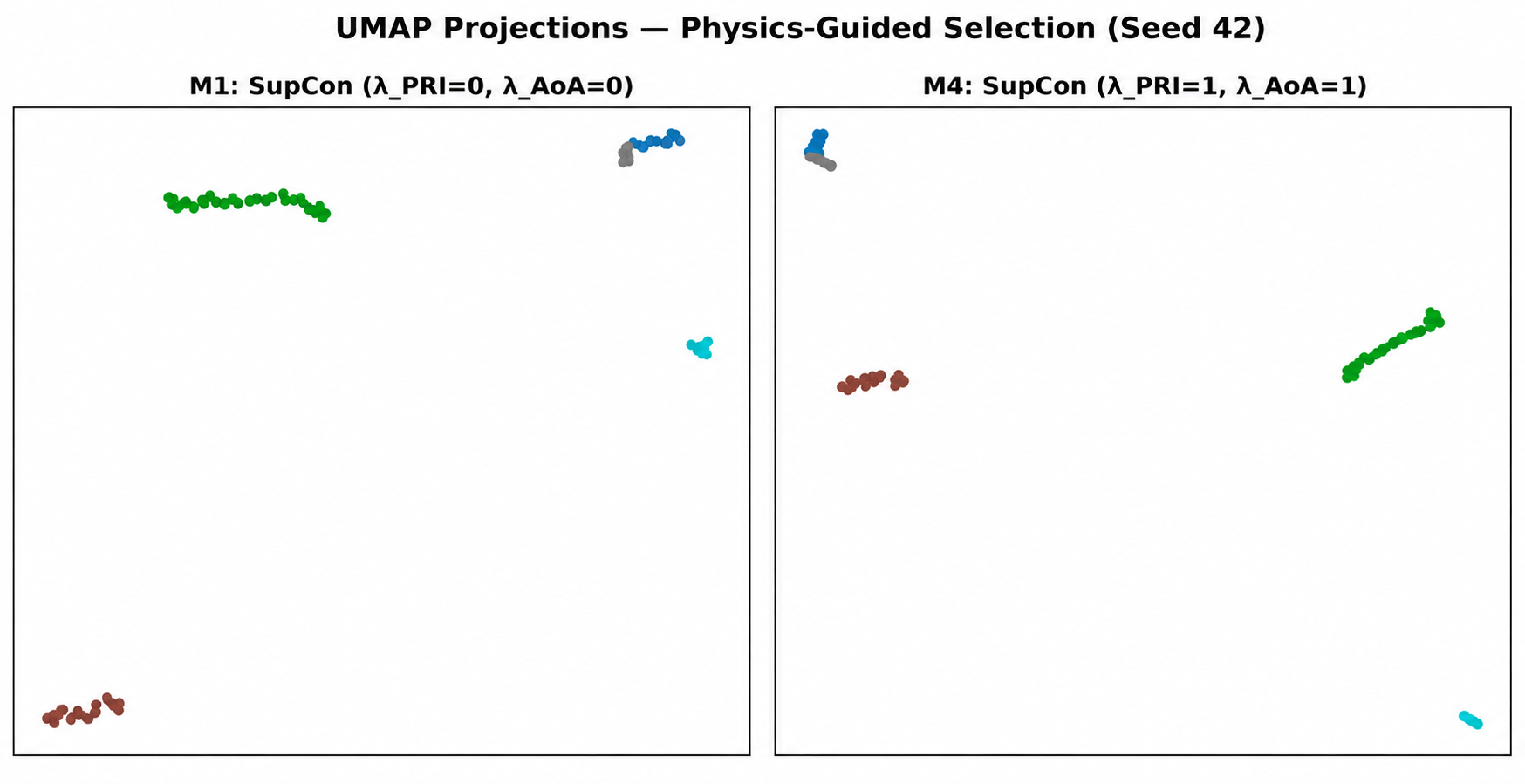}
	\caption{Empirical 2D UMAP projections of 128-dimensional encoder embeddings for M1 (left, SupCon only, unconstrained) vs. M4 (right, full PRI + AoA validation priors) on a 5-emitter synthetic test scenario (n\_neighbors=15, min\_dist=0.1)}
	\label{fig:UMAP_projections}
\end{figure*}

To qualitatively illustrate the impact of physical validation priors on the learned embedding space, we project the 128-dimensional encoder outputs to 2D using UMAP \citep{mcinnes2018umap} across representative test windows (Figure \ref{fig:UMAP_projections}).

\subsection{Statistical Significance (Wilcoxon Signed-Rank Test)}

As a supplementary pooled analysis, we apply a two-sided Wilcoxon signed-rank test to the M1 vs. M4 per-window Hungarian F1 differences across the full TSRD test set evaluated independently under each of the three random seeds, yielding 5,547 paired observations in total (3 seeds $\times$ 1,849 test windows). The test yields W\_stat = 6,742,879.0, p = 0.8649 (not significant at any conventional $\alpha$ threshold), rank-biserial correlation r\_B = 0.0027, and Cohen's d = 0.0222 (negligible effect size). This is consistent with the main finding that physics-guided AoA validation does not alter full-benchmark F1 (M1: 0.5813 vs. M4: 0.5790), while reducing cross-seed checkpoint-selection variance (M3/M4 SD = $\pm$0.0013 vs. M1 SD = $\pm$0.0063). We emphasise that contiguous overlapping evaluation windows (stride 128) exhibit local temporal correlation and do not constitute statistically independent observations; accordingly, the Wilcoxon result is reported as supplementary descriptive evidence only, not as a primary inferential claim.

\section{Conclusion and Limitations}

This paper presented a systematic empirical characterisation of continuous physical temporal representations and physics-informed domain priors in deep metric learning for open-world radar pulse de-interleaving. Our primary finding is that continuous ToA sinusoidal positional encoding yields a +133.4\% relative F1 improvement (0.8744 vs. 0.3747) and +87.5\% V-measure improvement (0.8484 vs. 0.4526) over discrete ordinal index encodings across three independent random seeds (Table IV). On the full open-world TSRD test set (1,849 windows), the unconstrained contrastive baseline M1 achieves the highest Hungarian F1 (0.5813) and CRR (0.8296). AoA-guided configurations M3 and M4 achieve F1 = 0.5790 with substantially lower cross-seed standard deviation ($\pm$0.0013 vs. $\pm$0.0063 for M1), indicating more stable checkpoint selection. On controlled synthetic streams, PRI-guided selection (M2) provides F1 advantages at low-to-moderate pulse loss (+0.0125 at 20\% loss) and across all clutter tiers (+0.0074 at 30\% clutter), while M1 matches or surpasses M2 at 30\% pulse loss and under combined harsh noise, demonstrating condition-dependent rather than universal physics-prior benefit. End-to-end latency profiling demonstrates sustained throughput exceeding 15,100 PDWs/s (16.85 ms total latency at W = 256), identifying downstream density clustering (HDBSCAN, 88.1\% of total pipeline latency) as the primary computational bottleneck for future GPU acceleration.

\subsection{Limitations and Future Work}

We explicitly identify four key limitations of the present study. First, evaluation is performed using the Turing Synthetic Radar Dataset (TSRD) and controlled generative models; while TSRD represents the leading unclassified benchmark, simulated PDW streams do not fully capture receiver hardware non-linearities, multipath reflections, and propagation anomalies encountered in real operational ELINT environments. Validating this framework on real-world intercept recordings represents a crucial next step. Second, current physics-guided validation criteria apply static weighting ($\lambda_{pri}, \lambda_{aoa}$); developing dynamic, noise-aware adaptive weighting that attenuates physical validation penalties upon detecting high clutter represents a high-priority research direction. Third, HDBSCAN clustering currently executes on CPU batches; implementing GPU-accelerated approximate nearest neighbour (ANN) clustering backends will further reduce latency for ultra-dense pulse streams. Finally, extending this architecture to online recursive streaming and multi-sensor ELINT fusion remains an exciting frontier.

\section*{Acknowledgements}

The authors acknowledge the use of large language model based tools for limited
language editing and formatting assistance. The authors take full responsibility
for the technical content, analysis, and conclusions presented in this work.

\bibliographystyle{unsrt}
\bibliography{references}

@inproceedings{gunn2025radar,
	title={Radar pulse deinterleaving with transformer based deep metric learning},
	author={Gunn, Edward and Hosford, Adam and Mannion, Daniel and Williams, Jarrod and Chhabra, Varun and Nockles, Victoria},
	booktitle={2025 IEEE International Radar Conference (RADAR)},
	pages={1--6},
	year={2025},
	organization={IEEE}
}

@inproceedings{gunn2026turing,
	title={The Turing Synthetic Radar Dataset: A Dataset for Pulse Deinterleaving},
	author={Gunn, Edward and Hosford, Adam and Jones, Robert and Zeitler, Leo and Groves, Ian and Nockles, Victoria},
	booktitle={2026 27th International Radar Symposium (IRS)},
	pages={78--83},
	year={2026},
	organization={IEEE}
}

@article{khosla2020supervised,
	title={Supervised contrastive learning},
	author={Khosla, Prannay and Teterwak, Piotr and Wang, Chen and Sarna, Aaron and Tian, Yonglong and Isola, Phillip and Maschinot, Aaron and Liu, Ce and Krishnan, Dilip},
	journal={Advances in neural information processing systems},
	volume={33},
	pages={18661--18673},
	year={2020}
}

@book{wiley2006elint,
	title={ELINT: The interception and analysis of radar signals},
	author={Wiley, Richard},
	year={2006},
	publisher={Artech}
}

@article{vaswani2017attention,
	title={Attention is all you need},
	author={Vaswani, Ashish and Shazeer, Noam and Parmar, Niki and Uszkoreit, Jakob and Jones, Llion and Gomez, Aidan N and Kaiser, {\L}ukasz and Polosukhin, Illia},
	journal={Advances in neural information processing systems},
	volume={30},
	year={2017}
}

@article{mcinnes2017hdbscan,
	title={hdbscan: Hierarchical density based clustering.},
	author={McInnes, Leland and Healy, John and Astels, Steve and others},
	journal={J. Open Source Softw.},
	volume={2},
	number={11},
	pages={205},
	year={2017}
}

@inproceedings{agnihotri2019effect,
	title={Effect of frequency on micro-Doppler signatures of a helicopter},
	author={Agnihotri, Vikas and Sabharwal, Munish and Goyal, Vinay},
	booktitle={2019 International Conference on Advances in Big Data, Computing and Data Communication Systems (icABCD)},
	pages={1--5},
	year={2019},
	organization={IEEE}
}

@inproceedings{agnihotri2019extraction,
	title={The extraction of key distinct features for identification and classification of helicopters using micro-doppler signatures},
	author={Agnihotri, Vikas and Sabharwal, Munish and Goyal, Vinay},
	booktitle={2019 IEEE Intl Conf on Dependable, Autonomic and Secure Computing, Intl Conf on Pervasive Intelligence and Computing, Intl Conf on Cloud and Big Data Computing, Intl Conf on Cyber Science and Technology Congress (DASC/PiCom/CBDCom/CyberSciTech)},
	pages={893--896},
	year={2019},
	organization={IEEE}
}

@article{agnihotri2020automatic,
	title={An automatic radar based aerial target recognition framework},
	author={Agnihotri, Vikas and Sabharwal, Munish},
	journal={Journal of Interdisciplinary Mathematics},
	volume={23},
	number={2},
	pages={321--333},
	year={2020},
	publisher={Taylor \& Francis}
}

@article{ge2019improved,
	title={Improved algorithm of radar pulse repetition interval deinterleaving based on pulse correlation},
	author={Ge, Zhipeng and Sun, Xian and Ren, Wenjuan and Chen, Wenbin and Xu, Guangluan},
	journal={IEEE access},
	volume={7},
	pages={30126--30134},
	year={2019},
	publisher={IEEE}
}

@inproceedings{mardia1989new,
	title={New techniques for the deinterleaving of repetitive sequences},
	author={Mardia, HK},
	booktitle={IEE Proceedings F (Radar and Signal Processing)},
	volume={136},
	number={4},
	pages={149--154},
	year={1989},
	organization={IET}
}

@article{liu2018classification,
	title={Classification, denoising, and deinterleaving of pulse streams with recurrent neural networks},
	author={Liu, Zhang-Meng and Philip, S Yu},
	journal={IEEE transactions on aerospace and electronic systems},
	volume={55},
	number={4},
	pages={1624--1639},
	year={2018},
	publisher={IEEE}
}

@article{chao2022radar,
	title={A radar signal deinterleaving method based on semantic segmentation with neural network},
	author={Chao, Wang and Liting, Sun and Zhangmeng, Liu and Zhitao, Huang},
	journal={IEEE Transactions on Signal Processing},
	volume={70},
	pages={5806--5821},
	year={2022},
	publisher={IEEE}
}

@inproceedings{chen2020simple,
	title={A simple framework for contrastive learning of visual representations},
	author={Chen, Ting and Kornblith, Simon and Norouzi, Mohammad and Hinton, Geoffrey},
	booktitle={International conference on machine learning},
	pages={1597--1607},
	year={2020},
	organization={PmLR}
}

@inproceedings{mcinnes2017accelerated,
	title={Accelerated hierarchical density based clustering},
	author={McInnes, Leland and Healy, John},
	booktitle={2017 IEEE international conference on data mining workshops (ICDMW)},
	pages={33--42},
	year={2017},
	organization={IEEE}
}

@article{raissi2019physics,
	title={Physics-informed neural networks: A deep learning framework for solving forward and inverse problems involving nonlinear partial differential equations},
	author={Raissi, Maziar and Perdikaris, Paris and Karniadakis, George E},
	journal={Journal of Computational physics},
	volume={378},
	pages={686--707},
	year={2019},
	publisher={Elsevier}
}

@article{mcinnes2018umap,
	title={Umap: Uniform manifold approximation and projection for dimension reduction},
	author={McInnes, Leland and Healy, John and Melville, James},
	journal={arXiv preprint arXiv:1802.03426},
	year={2018}
}

@article{genccol2017improvements,
	title={Improvements on deinterleaving of radar pulses in dynamically varying signal environments},
	author={Gen{\c{c}}ol, Kenan and Kara, Ali and At, Nuray},
	journal={Digital Signal Processing},
	volume={69},
	pages={86--93},
	year={2017},
	publisher={Elsevier}
}

@article{wang2022parametric,
	title={Parametric model-based deinterleaving of radar signals with non-ideal observations via maximum likelihood solution},
	author={Wang, Haiyu and Zhu, Mengtao and Fan, Ruozhou and Li, Yan},
	journal={IET radar, sonar \& navigation},
	volume={16},
	number={8},
	pages={1253--1268},
	year={2022},
	publisher={Wiley Online Library}
}

@article{wu2024efficient,
	title={Efficient FPGA implementation of convolutional neural networks and long short-term memory for radar emitter signal recognition},
	author={Wu, Bin and Wu, Xinyu and Li, Peng and Gao, Youbing and Si, Jiangbo and Al-Dhahir, Naofal},
	journal={Sensors},
	volume={24},
	number={3},
	pages={889},
	year={2024},
	publisher={MDPI}
}

@article{jiang2024radar,
	title={Radar pre-sorting algorithm based on autoencoder and lstm},
	author={Jiang, Yilin and Shi, Shaoxiong and Zhang, Fangyuan and Huang, Wuqi},
	journal={AEU-International Journal of Electronics and Communications},
	volume={187},
	pages={155535},
	year={2024},
	publisher={Elsevier}
}

\end{document}